# Dice in the Black Box: User Experiences with an Inscrutable Algorithm

Aaron Springer[1], Victoria Hollis[1], and Steve Whittaker[1]
University of California, Santa Cruz, 1156 High St., Santa Cruz, California, USA
{alspring, hollis, swhittak}@ucsc.edu

**Abstract**

We demonstrate that users may be prone to place an inordinate amount of trust in black box algorithms that are framed as intelligent. We deploy an algorithm that purportedly assesses the positivity and negativity of a users' writing emotional writing. In actuality, the algorithm responds in a random fashion. We qualitatively examine the paths to trust that users followed while testing the system. In light of the ease with which users may trust systems exhibiting "intelligent behavior" we recommend corrective approaches.

## Introduction

Hidden algorithms increasingly govern our lives. Hundreds of decisions pertaining to each of us are made each day. Google Now decides your route to work, Facebook chooses your most interesting friends for the day, and Netflix spins up your closely monitored "Understated Independent Dysfunctional-Family Dramas" recommendation category. These decisions made on our behalf directly influence our behavior, but researchers have only recently begun to examine the human aspects of these interactions.

We believe that complex algorithmic behavior should be explained to users. Prior research indicates that short textual explanations of algorithmic decision-making build trust with the algorithm's results (Kizilcec, 2016). However, explanation need not happen only in text. EmotiCal (Hollis et al., 2016) is a mobile mood regulation application that predicts users' future moods based upon previous activities and trends. The EmotiCal interface encourages users to actively explore how future mood predictions change based on new activities that they schedule. For example, they can see that adding an activity like "see a movie with a friend" boosts future mood, whereas "working alone" does not. This form of experiential explanation allows users to experiment with the system and draw obvious cause and effect inferences from their experimentation.

Other recent research suggests a rather different viewpoint. KnowMe (Warshaw et al., 2015) is a program that infers personality traits from a user's posts on social media based on Big Five personality theory. Users were simply told that the program analyzes their word use and correlates this with established personality surveys. We asked participants to evaluate KnowMe's outputs, in the form of a short paragraph describing their personality. Users were quick to defer to algorithmic judgment about their own personalities, stating that the algorithm is likely to have greater credibility than their own personal statements (e.g., "...At the end of the day, that's who the system says I am..."). This is particularly interesting because users ought to consider themselves to be true experts on their own personality. KnowMe users were also given the ability to correct the algorithm's judgment of their personality. Even when users disagreed with KnowMe's output, or when they felt that a negative portrayal was being presented, they still showed hesitancy to correct it. These findings indicate broad over-acceptance of algorithmic authority in the realm of personality judgments even when the algorithm fails to offer an explanation of the judgment. There are echoes here with research on persuasion where participants may overlook the content of an argument instead relying on the authority of the persuading agent (Petty et al., 1981).

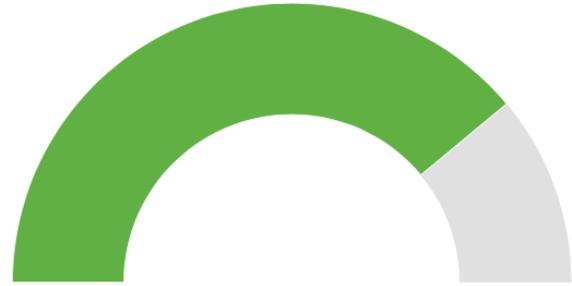

*Figure 1: The E-meter System*

We examine a new domain in the context of these recent paradoxical findings: mood. People often have difficulty interpreting their own emotions and predicting future mood states (Gilbert et al., 1998). In order to prompt individuals to more carefully consider their own moods and the reasons for these feelings, applications like Echo (Isaacs et al., 2013, Konrad et al., 2016a,b) have employed Technology Mediated Reflection (TMR). Echo asks users to reflect on previous memories and their relation to their mood, which in turn causes positive adaptions and higher wellbeing.

We propose that algorithmic evaluation of people's writing can be used to facilitate TMR. Using data from previous studies like EmotiCal and Echo, we have built highly accurate models that predict user mood from the journal entries they write. Would displaying feedback from this algorithm in real-time cause people to carefully consider their writing and reflect on how the events they write about affected them?

Asking this question led us to consider the relationship between algorithm accuracy and user response to the algorithm's presentation and context. In order to draw a baseline and disaggregate these factors we removed algorithm accuracy from the equation. Rather than having users respond to a carefully constructed model of mood, we asked them to respond to random responses framed as intelligence. Participants were presented with a webpage, and asked to write about their personal experiences over the last 3 days. They were told that an intelligent algorithm, the *E-meter,* was monitoring their writing and interpreting their feelings. The random feedback was presented in a "mood meter". We then asked them a series of questions about algorithmic accuracy, plausibility, and acceptability.

## Methods

### E-meter System

The E-meter system (Figure 1) presented participants with a webpage that contained a short set of instructions, a text box, and a gauge that changed in real time as they wrote. Underneath the graphic was a short explanation: "The graphic above displays the output from an algorithm that assesses the positivity/negativity of your writing as you answer the question below." The user was prompted to pick 3 events that had emotionally affected them in the past 3 days and then write a paragraph about how and why each event had affected them.

As the participants answered the prompt, the E-meter reacted by fluctuating in sync with the participant's writing. Given that our primary interest was user's responses to the system we programmed the E-meter trend either positive or negative rather than actually evaluate the user's writing. We expected that this would produce expectation violation for some users, which previous research has indicated plays a major role in responses to algorithmic interfaces (Kizilcec, 2016). The E-meter responded randomly with a positive or negative bias to each word over 4 characters. This gave the illusion that the algorithm ignored many uninformative words such as "is", "the", and "of". The positive and negative bias was small so that the E-meter graphic would often show small trends in either direction but overall the trend would slowly grow positive or negative. We hypothesized that this set of behaviors would lead to believable system where users could not easily tell that the algorithm was not actually "assess[ing] the positivity/negativity of your writing".

### Participants

Participants were recruited in 2 separate stages. Four participants were recruited initially from the authors' connections. These participants were primarily used to fix the initial system presentation, wording, and instructions. Between each of the 4 participants the system was modified to better address the concerns and flaws that were exposed by the previous participant.

The second stage consisted of 8 participants recruited from the /r/SampleSize community on Reddit. While demographic information was not gathered, it is likely that these participants mirrored the Reddit population as a whole. Reddit users tend to be younger, male biased, and more technologically savvy than the general population (Barthel et al., 2016). These users were asked to carry out the procedure on a volunteer basis.

### Materials and Methodology

After users wrote about 3 emotionally charged events, we asked a number of questions. These primarily focused on users' judgments of the accuracy of the E-meter and its affects on their writing.

The responses were first coded using an open coding method; this created 13 initial codes. Thematic coding was then used to group the open codes into 7 major themes. We focus primarily on 2 of these themes: user testing of the algorithm and user internal states compared to user represented state.

## Results

While a sample of 12 participants is far too small to draw an accurate quantitative representation of the population, we report accuracy to satisfy curiosity. Although we are aware of the limited nature of our data, we also present a qualitative analysis of responses, suggesting behaviors and attitudes that give rise to important design considerations.
Users rated the E-meter accuracy on a 7-point scale with possible responses ranging from 1 (very inaccurate) to 7

(very accurate). The median report for users' assessment of the E-meter's accuracy was 5 (see Figure 2), which indicates that the E-meter was "Slightly Accurate".

Many users tested the algorithm as they were writing. Some did so by consciously manipulating the application

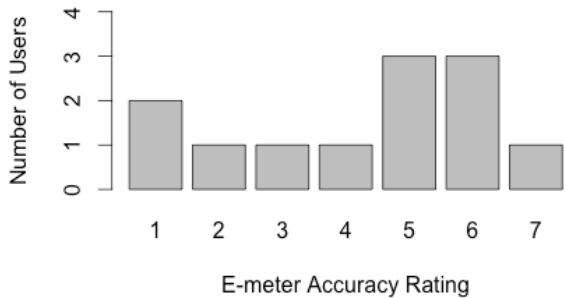

and watching the changes in the E-meter according to each word they wrote:

> "I used 'father' and it went up, obviously thinking family = nice. I used 'trust' and it went up, obviously thinking trust = good." – Participant 9

Others explicitly changed their writing in order to see how the E-meter responded:

> "Its lack of sensitivity to my adding a couple more "positive" words at the end of my third paragraph after the rest of my writing had been strongly negative was also somewhat impressive." - Participant 3

> "I was messing with it to see if more positive words would give a more positive rating." – Participant 11

Even though many users explicitly tested the random E-meter, they still found it slightly accurate. We attribute this to a combination of confirmation bias and peripheral routes to persuasion (Petty et al., 1981). Even Participant 9, who demonstrated to their own satisfaction that the E-meter was wrong, explicitly talks about how their expectations of the algorithm's performance were confirmed. Perhaps more perniciously, participants were highly likely to disregard opposing evidence, like the E-meter going down when they wrote something happy, due to the framing of the system as AI-like. In other writing we see that many users make explicit excuses for why the E-meter's output did not match their mood exactly. Participant 5 wrote:

> "I did not give it enough to work with"

The clearest evidence of confirmation bias is revealed in how Participant 10 wrote:

> "It was somewhat inaccurate as one of my days items was negative. My other 2 were very positive so - yeah - it was rather accurate."

We also saw some evidence for algorithmic omniscience. Three participants wrote in ways that suggested that they believed something deeper might be happening with the algorithm, almost as if it could see through the text into how they were really feeling. They felt that the way they represented themselves in their writing was not always an accurate mirror of how they truly felt. In one example, Participant 4 rated their writing's mood a 3 and the E-meter "assessed" the writing's mood as a 1. Participant 4 then wrote (emphasis ours):

> ***"While this week was mostly an emotionally chal-***

Figure 2: E-meter User Accuracy Ratings

> ***lenging one for me,*** *I mostly listed minor issues through the week as well as some positive ones.* ***Strictly from the writing,*** *it should have probably read closer to a neutral or only slightly negative state."*

In another similar example Participant 8 rated the accuracy of the E-meter as "Accurate" and then wrote (emphasis ours):

> ***"I'm not in a great mood,*** *but it was all the way bad and One of my three memories was plesant [sic]"*

These, among others, illustrate that participants engaged in the exercise with two different perspectives on how they were representing themselves. They write about an internal self ("I'm not in a great mood") and also a represented self ("Strictly from the writing"). The example data presented here suggests that they viewed the algorithm as almost omniscient, seeing through their represented selves into their internal selves. This has potentially troubling consequences for people's knowledge of algorithms as well as for how we present their results.

## Discussion

We are aware that we have conducted a small pilot. Nevertheless, combined with consistent results from other recent studies (Warshaw et al., 2015, Costa et al., 2016), important issues arise concerning algorithm perception and authority. Nearly twice as many users rated our completely random system as accurate rather than inaccurate. This is especially critical given the context in which the algorithm is operating. As in the Warshaw et al. (2015) personality study, the participants themselves are the true experts of their own mood. Nevertheless, participants are willing to consider that an algorithm can know their mood as well or better than themselves. This demonstrates the trust and authority that we imbue algorithms with in our daily lives.

The first route through which participants gained trust in the system was through their own active testing of the E-

meter. Users typed words that they associated with positive or negative moods and saw how the meter responded, often concluding that it was accurate. From an objective standpoint this seems absurd. The E-meter is literally responding randomly to words over 4 characters and a short experiment should be enough to dispel the AI myth.

However, this underestimates the biases that we all hold. We are programmed to look for patterns in randomness (Foster and Kokko, 2009). Combine this predilection for pattern seeking with peripheral persuasive routes and considering a random algorithm framed as AI to be accurate begins to make sense. Users are essentially seeking to confirm what they have already been told rather than evaluating the system on its merits. Without explanation, users are forced to rely on peripheral routes including the perceived prestige and trustworthiness of the company deploying the algorithms (Petty et al., 1981).

The second major route for participants to trust the E-meter was through a belief in the omnipotence of the algorithm. This directly echoes results of the KnowMe system where users felt that its accuracy was "eerie" given the small amount of information it used (Warshaw et al., 2015). What is particularly troubling for the E-meter is that this happened with random output. This suggests that some users may be overly deferential to an algorithm's evaluation, even for areas in which they have expertise.

Both of these routes result in users placing far too much trust in algorithmic decisions. The consequences of this may already be realized in the use of the Tesla Autopilot system. Users have ignored recommendations and the explicit warnings from the system to keep hands on the steering wheel at all times and to be ready for a manual takeover and this has resulted in 2 confirmed crashes, including one fatality (Tesla, 2016; Isidore and Sung, 2016).

In addition, more research is needed to address the potential psychological consequences of trusting algorithm feedback. For example, recent findings have suggested that false feedback on stress states can impact the perceived stress of a difficult task (Costa et al., 2016). Rather than only changing thoughts, incorrect inferences can fundamentally influence emotional states and behavior. We find this troubling given the 165,000 health related applications available from Google Play and iTunes. Our results suggest that users may be over willing to credit these applications, with possible negative consequences for their health.

Our results suggest that great care must be taken in designing applications that "explain" algorithms. Users' criteria for explanation may be much less stringent that we have previously thought. We recommend further research into experiential explanation and how to use this to calibrate expectations.

As initial steps we call for best practices for explanation. Best practices for explanation entails first following the example set in (Kizilcec 2016) of short explanations and then, one step further, providing access to resources that explain the functioning in more detail.

Long term we believe that learning algorithms making impactful decisions must be audited by a regulatory third party. We envision this regulator filling a similar role as the Weights and Measures Division in the USA. This would make sure the veneer of intelligence is not simply snake oil underneath.

# References

Barthel, M., Stocking, G., Holcomb, J., & Mitchell, A. (2016, February 25). 1. Reddit news users more likely to be male, young and digital in their news preferences. Retrieved from http://www.journalism.org/2016/02/25/reddit-news-users-more-likely-to-be-male-young-and-digital-in-their-news-preferences/

Costa, J., Adams, A. T., Jung, M. F., Guimbetière, F., & Choudhury, T. (2016). EmotionCheck: leveraging bodily signals and false feedback to regulate our emotions (pp. 758–769). ACM Press. https://doi.org/10.1145/2971648.2971752

Foster, K. R., & Kokko, H. (2009). The evolution of superstitious and superstition-like behaviour. *Proceedings of the Royal Society B: Biological Sciences*, *276*(1654), 31–37. https://doi.org/10.1098/rspb.2008.0981

Gilbert, D. T., Pinel, E. C., Wilson, T. D., Blumberg, S. J., & Wheatley, T. P. (1998). Immune neglect: a source of durability bias in affective forecasting. *Journal of Personality and Social Psychology*, *75*(3), 617.

Hollis, V., Konrad, A., Springer, A., Antoun, C., Antoun, M., Martin, R., & Whittaker, S. (2016). What Does All This Data Mean For My Future Mood? Actionable Analytics & Targeted Reflection for Emotional Well-Being. Under review.

Isidore, C., & Sung, G. (2016, July 12). Driver in Tesla Autopilot accident would buy another Tesla. Retrieved October 27, 2016, from http://money.cnn.com/2016/07/12/technology/tesla-autopilot-accident/index.html

Kizilcec, R. F. (2016). How Much Information?: Effects of Transparency on Trust in an Algorithmic Interface (pp. 2390–2395). ACM Press. https://doi.org/10.1145/2858036.2858402

Konrad, A., Tucker, S., Crane, J., & Whittaker, S. (2016). Technology and Reflection: Mood and Memory Mechanisms for Well-Being. *Psychology of Well-Being*, *6*(1). https://doi.org/10.1186/s13612-016-0045-3

Petty, R. E., Cacioppo, J. T., & Goldman, R. (1981). Personal involvement as a determinant of argument-based persuasion. *Journal of Personality and Social Psychology*, *41*(5), 847.

Tesla. (2016, June 30). A Tragic Loss. Retrieved October 27, 2016, from https://www.tesla.com/blog/tragic-loss

Warshaw, J., Matthews, T., Whittaker, S., Kau, C., Bengualid, M., & Smith, B. A. (2015). Can an Algorithm Know the "Real You"?: Understanding People's Reactions to Hyper-personal Analytics Systems (pp. 797–806). ACM Press. https://doi.org/10.1145/2702123.2702274